\documentclass[aps,pre,showpacs,noshowkeys,amsmath,amssymb,amsfonts,superscriptaddress,longbibliography,reprint]{revtex4-1}
\usepackage[english]{babel}

\usepackage{graphicx}
\usepackage{bm}
\usepackage{physics}
\usepackage{mathtools}
\usepackage{gensymb}
\usepackage{array}
\newcolumntype{C}[1]{>{\centering\let\newline\\\arraybackslash\hspace{0pt}}m{#1}}

\setcitestyle{super}
\usepackage{caption}
\usepackage{subcaption}
\usepackage{sidecap}

\DeclareCaptionLabelSeparator{bar}{~\rule[-0.4ex]{0.2ex}{1em}~}
\DeclareCaptionLabelFormat{subfor}{(\textbf{#2})}
\usepackage[dvipsnames]{xcolor}
\definecolor{UBcolor}{HTML}{007CC1}
\usepackage[colorlinks=true,pdfnewwindow=true,linkcolor=UBcolor,citecolor=UBcolor,urlcolor=UBcolor,breaklinks=true,linktocpage]{hyperref}
\usepackage[all]{hypcap}
\usepackage[nameinlink,capitalise]{cleveref}
\crefname{SI section}{SI Section}{SI Sections}
\Crefname{SI section}{SI Section}{SI Sections}
\begin{document}

\title{Chemotactic smoothing of collective migration}

\author{Tapomoy Bhattacharjee}
\altaffiliation{These authors contributed equally to this work.}
\affiliation{The Andlinger Center for Energy and the Environment, Princeton University, Princeton, NJ, 08544, USA}

\author{Daniel B. Amchin}
\altaffiliation{These authors contributed equally to this work.}
\affiliation{Department of Chemical and Biological Engineering, Princeton University, Princeton, NJ, 08544, USA}

\author{Ricard Alert}
\altaffiliation{These authors contributed equally to this work.}
\affiliation{Lewis-Sigler Institute for Integrative Genomics, Princeton University, Princeton, NJ 08544, USA}
\affiliation{Princeton Center for Theoretical Science, Princeton University, Princeton, NJ 08544, USA}

\author{J. A. Ott}
\affiliation{Department of Chemical and Biological Engineering, Princeton University, Princeton, NJ, 08544, USA}

\author{Sujit S. Datta}
\email{ssdatta@princeton.edu}
\affiliation{Department of Chemical and Biological Engineering, Princeton University, Princeton, NJ, 08544, USA}

\date{\today}

\begin{abstract}
Collective migration---the directed, coordinated motion of many self-propelled agents---is a fascinating emergent behavior exhibited by active matter that has key functional implications for biological systems. Extensive studies have elucidated the different ways in which this phenomenon may arise. Nevertheless, how collective migration can persist when a population is confronted with perturbations, which inevitably arise in complex settings, is poorly understood. Here, by combining experiments and simulations, we describe a mechanism by which collectively migrating populations smooth out large-scale perturbations in their overall morphology, enabling their constituents to continue to migrate together. We focus on the canonical example of chemotactic migration of \textit{Escherichia coli}, in which fronts of cells move \textit{via} directed motion, or chemotaxis, in response to a self-generated nutrient gradient. We identify two distinct modes in which chemotaxis influences the morphology of the population: cells in different locations along a front migrate at different velocities due to spatial variations in (i) the local nutrient gradient and in (ii) the ability of cells to sense and respond to the local nutrient gradient. While the first mode is destabilizing, the second mode is stabilizing and dominates, ultimately driving smoothing of the overall population and enabling continued collective migration. This process is autonomous, arising without any external intervention; instead, it is a population-scale consequence of the manner in which individual cells transduce external signals. Our findings thus provide insights to predict, and potentially control, the collective migration and morphology of cell populations and diverse other forms of active matter.
\end{abstract}

\maketitle

The flocking of birds, schooling of fish, herding of animals, and procession of human crowds are all familiar examples of collective migration. This phenomenon also manifests at smaller scales, such as in populations of cells and dispersions of synthetic self-propelled particles. In addition to being a fascinating example of emergent behavior, collective migration can be critically important---enabling populations to follow cues that would be undetectable to isolated individuals \cite{Camley2018}, escape from harmful conditions and colonize new terrain \cite{Cremer2019}, and coexist \cite{Gude2020a}. Thus, diverse studies have sought to understand the mechanisms by which collective migration can arise.

Less well understood, however, is how collective migration persists after a population is confronted with perturbations. These can be external, stemming from heterogeneities in the environment \cite{Sandor,Morin2017,Wong2014,Chepizhko2013,Chepizhko2013a,Chepizhko2015,Toner2018,Maitra2020a}, or internal, stemming from differences in the behavior of individuals \cite{Yllanes2017,Bera2020,Alirezaeizanjani2020}. Mechanisms by which such perturbations can \textit{disrupt} collective migration are well documented. Indeed, in some cases, perturbations can abolish coordinated motion throughout the population entirely \cite{Sandor,Morin2017,Yllanes2017,Bera2020,Chepizhko2013,Chepizhko2013a,Chepizhko2015,Toner2018}. In other cases, perturbations couple to the active motion of the population to destabilize its leading edge, producing large-scale disruptions to its morphology \cite{Wong2014,Alert2020,Alert2019,Driscoll2017a,Doostmohammadi2016,Williamson2018,saverio}. Indeed, for one of the simplest cases of collective migration---\textit{via} chemotaxis, the biased motion of cells up a chemical gradient---morphological instabilities can occur due to the disruptive influence of hydrodynamic \cite{koch,kela1,kela2} or chemical-mediated \cite{BenAmar2016,BenAmar2016b,funaki,brenner,mimura,stark} interactions between cells. By contrast, mechanisms by which migrating populations can \textit{withstand} perturbations have scarcely been examined.

\begin{figure*}
\centering
\includegraphics[width=0.8\textwidth]{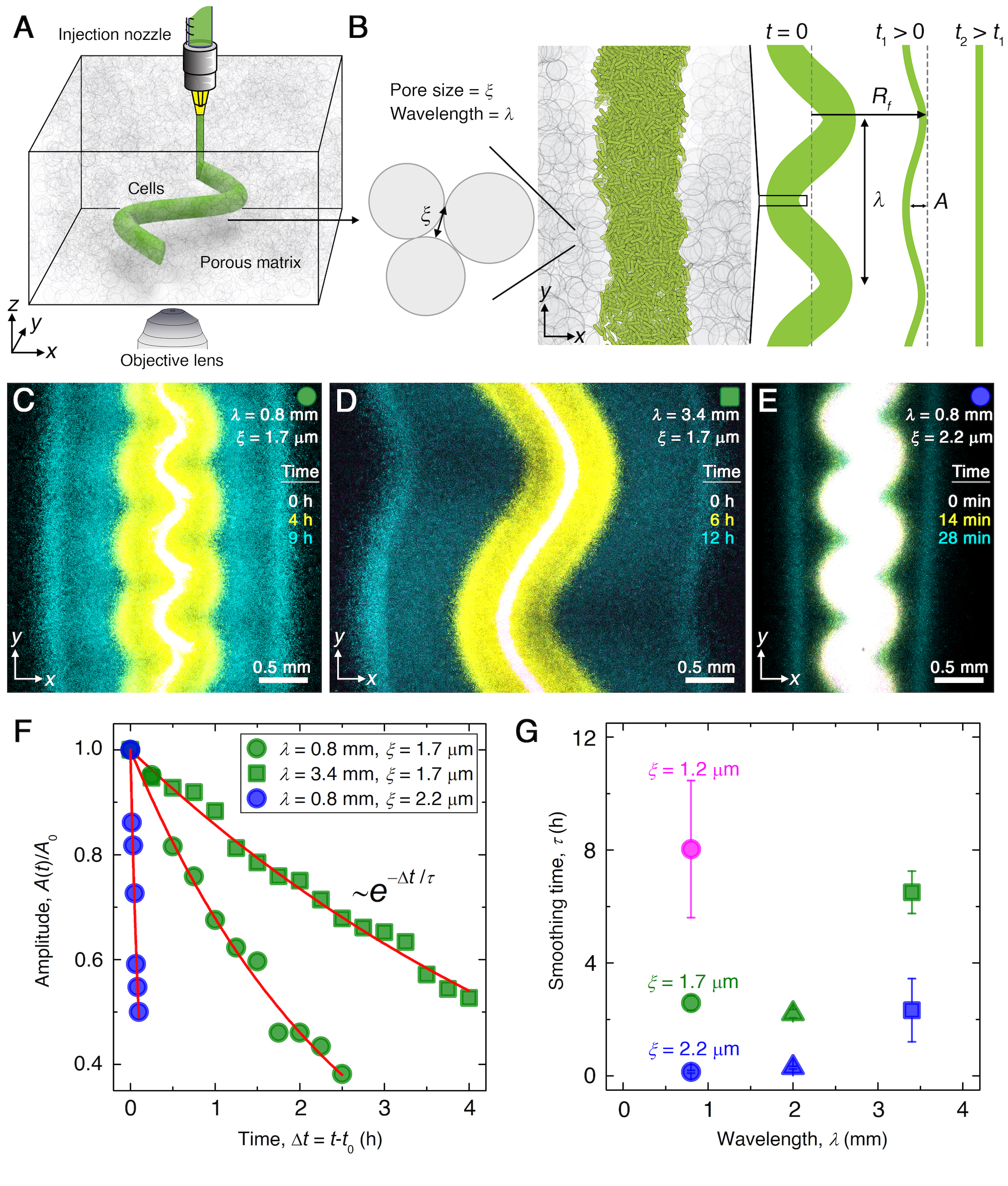}
  {\phantomsubcaption\label{Fig1a}}
  {\phantomsubcaption\label{Fig1b}}
  {\phantomsubcaption\label{Fig1c}}
  {\phantomsubcaption\label{Fig1d}}
  {\phantomsubcaption\label{Fig1e}}
  {\phantomsubcaption\label{Fig1f}}
  {\phantomsubcaption\label{Fig1g}}
\caption{\textbf{Experiments reveal that migrating \textit{E. coli} populations autonomously smooth large-scale morphological perturbations.} \subref*{Fig1a} Schematic of an undulated population (green cylinder) 3D-printed within a porous medium made of jammed hydrogel particles (gray). Each undulated cylinder requires $\sim10$ s to print, two orders of magnitude shorter than the duration between successive 3D confocal image stacks, $\sim10$ min. The surrounding medium fluidizes as cells are injected into the pore space, and then rapidly re-jams around the dense-packed cells. \subref*{Fig1b} Two-dimensional $xy$ slice through the mid-plane of the population. The starting morphology of the 3D-printed population has undulation wavelength $\lambda$ and amplitude $A_{0}$, as defined by the undulated path traced out by the injection nozzle. The cells subsequently swim through the pores between hydrogel particles, with mean pore size $\xi$. The population thereby migrates outward in a coherent front that eventually smooths; we track the radial position of the leading edge of the front $R_f$ and the undulation amplitude $A$ over time $t$. \subref*{Fig1c}-\subref*{Fig1e} Bottom-up ($xy$ plane) projections of cellular fluorescence intensity measured using 3D confocal image stacks. Images show sections of three initially undulated populations in three different porous media, each at three different times (superimposed white, yellow, cyan), as the cells migrate radially outward. A pixel corresponds to $\sim1$ cell, and the images only show a magnified view of the overall population. Panels \subref*{Fig1c}-\subref*{Fig1d} demonstrate the influence of varying the undulation wavelength, keeping the mean pore size the same; increasing $\lambda$ slows smoothing. Panels \subref*{Fig1c} and \subref*{Fig1e} demonstrate the influence of varying the pore size, keeping the undulation wavelength the same; increasing $\xi$ hastens smoothing. \subref*{Fig1f} For each experiment shown in \subref*{Fig1c}-\subref*{Fig1e}, the undulation amplitude $A$, normalized by its initial value $A_{0}$, decays exponentially with the time $\Delta t$ elapsed from the initiation of smoothing at $t=t_0$. Fitting the data (symbols) with an exponential decay (red lines) yields the smoothing time $\tau$ for each experiment. \subref*{Fig1g} Smoothing time $\tau$ measured in experiments increases with increasing undulation wavelength $\lambda$ and decreasing medium mean pore size $\xi$, which enables cells to migrate more easily. Error bars reflect the uncertainty in determining the initiation time $t_0$ from the exponential fit of the data.}
\label{fig:fig1}
 \end{figure*}

Here, we demonstrate a mechanism by which collectively migrating populations of \textit{E. coli} autonomously smooth out large-scale perturbations in their overall morphology. We show that chemotaxis in response to a self-generated nutrient gradient provides both the driving force for collective migration and the primary smoothing mechanism for these bacterial populations. Using experiments on 3D printed populations with defined morphologies, we characterize the dependence of this active smoothing on the wavelength of the perturbation and on the ability of cells to migrate. Furthermore, using continuum simulations, we show that the limited ability of cells to sense and respond to a nutrient gradient causes them to migrate at different velocities at different positions along a front---ultimately driving smoothing of the overall population and enabling continued collective migration. Our work thus reveals how cellular signal transduction enables a population to withstand large-scale perturbations, and provides a framework to predict and control chemotactic smoothing for active matter in general.

\section*{Results}

\subsection*{Chemotactic smoothing is regulated by\\perturbation wavelength and cellular motility} To experimentally investigate the collective migration of \textit{E. coli} populations, we confine them within porous media of tunable properties \cite{bhattacharjee2019bacterial,bhattacharjee2019confinement,Bhattacharjee2020}, as schematized in \cref{Fig1a,Fig1b} and detailed in the \hyperref[methods]{Materials and Methods}. The media are composed of hydrogel particles that are swollen in a defined rich liquid medium with \textit{L}-serine as the primary nutrient and chemoattractant. We enclose the particles at prescribed jammed packing fractions in transparent chambers. Because the hydrogel is highly swollen, it is freely permeable to oxygen and nutrient. However, while the particles do not hinder exposure of bacteria to these chemical signals, the cells cannot penetrate the individual particles, and are instead forced to swim through the interparticle pores (\cref{Fig1b}). Varying the hydrogel particle packing density thus enables us to tune pore size and thereby modulate cellular migration without altering the nutrient field \cite{bhattacharjee2019bacterial,bhattacharjee2019confinement,Bhattacharjee2020}. Specifically, we vary the mean pore size $\xi$ between $1.2~\mu$m and $2.2~\mu$m, causing cellular migration through the pore space to be more and less hindered, respectively, without deforming the solid matrix \cite{bhattacharjee2019bacterial}. Moreover, the packings are transparent, enabling the morphologies of the migrating populations to be tracked in the $xy$ plane using confocal fluorescence microscopy (\cref{Fig1a}); to this end, we use cells that constitutively express green fluorescent protein throughout their cytoplasm. 

A key feature of the hydrogel packings is that they are yield-stress solids; thus, an injection micronozzle can move along a prescribed path inside each medium by locally rearranging the particles, gently extruding densely-packed cells into the interstitial space (\cref{Fig1a,Fig1b}). The particles then rapidly re-densify around the newly-introduced cells, re-forming a jammed solid matrix that supports the cells in place with minimal alteration to the overall pore structure \cite{bhattacharjee2015,bhattacharjee2016,bhattacharjee2018}. This approach is therefore a form of 3D printing that enables the initial morphology of each bacterial population to be defined within the porous medium. The cells subsequently swim through the pores between particles, migrating outward through the pore space. For example, as we showed previously \cite{Bhattacharjee2020}, cells of \textit{E. coli} initially 3D printed in densely-packed straight cylinders collectively migrate radially outward in flat, coherent fronts. These fronts form and propagate \textit{via} chemotaxis: the cells continually consume surrounding nutrient, generating a local gradient that they in turn bias their motion along \cite{adler1966effect,Cremer2019,fu2018spatial,saragosti2011directional}. As this front of cells migrates, it propagates the local nutrient gradient with it through continued consumption, thereby sustaining collective migration. In the absence of nutrient, migrating fronts do not form at all \cite{Bhattacharjee2020}.

To test how perturbations in the overall morphology of the population influence its subsequent migration, we 3D print densely-packed \textit{E. coli} in 1 cm-long cylinders with spatially-periodic undulations as perturbations prescribed along the $x$ direction (\cref{Fig1b}). Each population is embedded deep within a defined porous medium; an initial population morphology is schematized at time $t=0$ in \cref{Fig1b}, with the undulation wavelength and amplitude denoted by $\lambda$ and $A$, respectively. An experimental realization with $A(t=0)\approx 300~\mu$m, $\lambda\approx0.8$ mm, and $\xi=1.7~\mu$m is shown in white in \cref{Fig1c}, which shows an $xy$ cross section through the midplane of the population. After 3D printing, the outer periphery of the population spreads slowly, hindered by cell-cell collisions in the pore space, as the population establishes a steep gradient of nutrient through consumption \cite{Bhattacharjee2020}. Then, this periphery spontaneously organizes into a $\sim300~\mu$m-wide front of cells that collectively migrates outward (yellow in \cref{Fig1c}). The undulated morphology of this front initially retains that of the initial population. Strikingly, however, the front autonomously smooths out these large-scale undulations as it continues to propagate (\hyperref[movies]{Movie S1}). We characterize this behavior by tracking the decay of the undulation amplitude, normalized by its initial value $A_{0}\equiv A(\Delta t=0)$, as a function of time elapsed from the initiation of smoothing, $\Delta t$ (green circles in \cref{Fig1f}). The normalized amplitude decays exponentially (red line in \cref{Fig1f}), with a characteristic time scale $\tau\approx2.5$ h, and the population eventually continues to migrate as a completely flat front (cyan in \cref{Fig1c}). 

We observe similar behavior when the wavelength $\lambda$ is increased to $3.4$ mm (\cref{Fig1d}, \hyperref[movies]{Movie S2}) or when the pore size $\xi$ is increased to $2.2~\mu$m (\cref{Fig1e}, \hyperref[movies]{Movie S3}); however, the dynamics of front smoothing are altered in both cases. Specifically, increasing the undulation wavelength slows smoothing, increasing $\tau$ by a factor of $\approx3$ to reach $\tau\approx6.5$ h (green squares in \cref{Fig1f}). Conversely, increasing the pore size---which enables cells to migrate through the pore space more easily---greatly hastens smoothing, decreasing $\tau$ by more than a factor of $\approx10$ to become $\tau\approx0.2$ h (blue circles in \cref{Fig1f}). This behavior is consistent across multiple experiments with varying $\lambda$ and $\xi$, as summarized in \cref{Fig1g}. Our experiments thus indicate that the smoothing of collective migration is regulated by both the undulation wavelength and the ease with which cells migrate.

\subsection*{A continuum model of chemotactic migration recapitulates\\the spatio-temporal features of smoothing} To gain further insight into the processes underlying smoothing, we use the classic Keller–Segel model of chemotactic migration \cite{lauffenburger1991quantitative,Keller1971a} to investigate the dynamics of undulated populations. Variants of this model can successfully capture the key features of chemotactic migration of flat \textit{E. coli} fronts in bulk liquid \cite{Keller1971a,fu2018spatial} and in porous media \cite{Bhattacharjee2020}; we therefore hypothesize that it can also help identify the essential physics of smoothing.

\begin{figure*}
\centering
\includegraphics[width=0.75\textwidth]{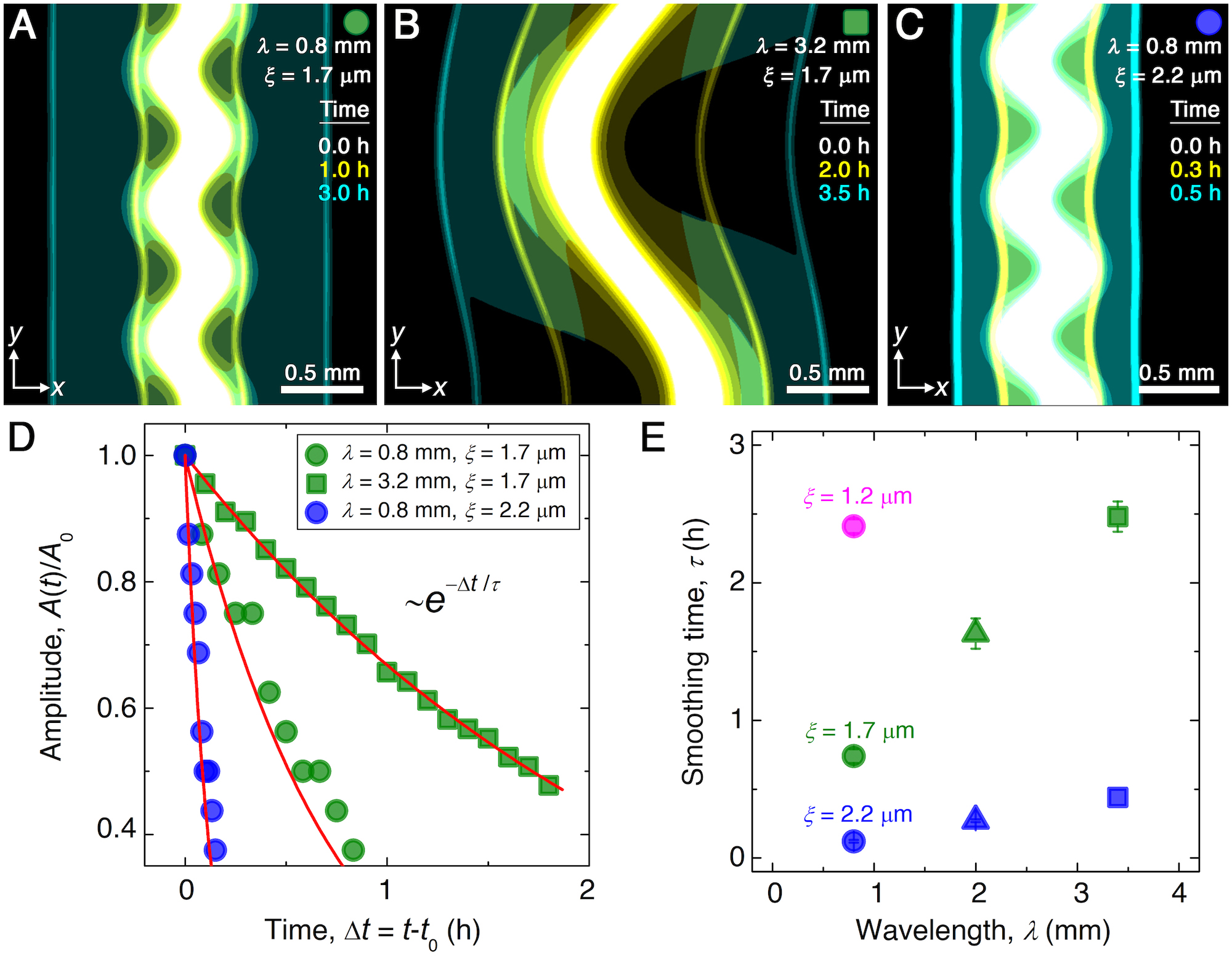}
  {\phantomsubcaption\label{Fig2a}}
  {\phantomsubcaption\label{Fig2b}}
  {\phantomsubcaption\label{Fig2c}}
  {\phantomsubcaption\label{Fig2d}}
  {\phantomsubcaption\label{Fig2e}}
\caption{\textbf{Continuum model captures the essential features of the smoothing of migrating bacterial populations.} \subref*{Fig2a}-\subref*{Fig2c} Simulations corresponding to experiments reported in \cref{Fig1c,Fig1e}, respectively, performed by numerically solving \cref{eq nutrient,eq bacteria-main-withgrowth} in two dimensions ($xy$ plane). Images show the calculated cellular signal (details in \hyperref[methods]{Materials and Methods}) for three initially undulated populations in three different porous media, each at three different times (superimposed white, yellow, cyan), as the cells migrate outward. Panels \subref*{Fig2a}-\subref*{Fig2b} demonstrate the influence of varying the undulation wavelength, keeping the mean pore size the same; as in the experiments, increasing $\lambda$ slows smoothing. Panels \subref*{Fig2a} and \subref*{Fig2c} demonstrate the influence of varying the pore size, keeping the undulation wavelength the same; as in the experiments, increasing $\xi$, incorporated in the model by using larger values of the diffusion and chemotactic coefficients as obtained directly from experiments, hastens smoothing. \subref*{Fig2d} For each simulation shown in \subref*{Fig2a}-\subref*{Fig2c}, the undulation amplitude $A$,normalized by its initial value $A_{0}$, decays exponentially with the time $\Delta t$ elapsed from the initiation of smoothing at $t=t_0$ as in the experiments. Fitting the data (symbols) with an exponential decay (red lines) again yields the smoothing time $\tau$ for each simulation. \subref*{Fig2e} Smoothing time $\tau$ obtained from the simulations increases with increasing undulation wavelength $\lambda$ and decreasing medium mean pore size $\xi$, as in the experiments. Error bars reflect the uncertainty in determining the initiation time $t_0$ from the exponential fit of the data.}
\label{fig:fig2}
\end{figure*}

To this end, we consider a two-dimensional (2D) representation of the population in the $xy$ plane for simplicity, with $\bm{r}\equiv(x,y)$, and model the evolution of the nutrient concentration $c(\bm{r},t)$ and number density of bacteria $b(\bm{r},t)$ using the coupled equations: 
\begin{equation} \label{eq nutrient}
\partial_t c = D_{\text{c}}\nabla^{2}c-b\kappa g(c),
\end{equation}
\begin{equation} \label{eq bacteria-main-withgrowth}
\partial_t b = -\bm{\nabla}\cdot \bm{J}_{\text{b}}+b\gamma g(c),\qquad \bm{J}_{\text{b}} = -D_{\text{b}}\bm{\nabla} b +  b \chi \bm{\nabla} f(c).
\end{equation}
\Cref{eq nutrient} relates changes in $c$ to nutrient diffusion and consumption by the bacteria; $D_{\text{c}}$ is the nutrient diffusion coefficient, $\kappa$ is the maximal consumption rate per cell, and $g(c)=c/\left(c+c_{\text{1/2}}\right)$ describes the influence of nutrient
availability relative to the characteristic concentration $c_{\text{1/2}}$ through Michaelis-Menten kinetics. \Cref{eq bacteria-main-withgrowth} relates changes in $b$ to the bacterial flux $\bm{J}_{\text{b}}$, which arises from their undirected and directed motion, and net cell proliferation with a maximal rate $\gamma$. In the absence of a nutrient gradient, bacteria move in an unbiased random walk \cite{Bergbook}; thus, undirected motion is diffusive over large length and time scales, with an effective diffusion coefficient $D_\text{b}$ whose value depends on both cellular activity and confinement in the pore space \cite{bhattacharjee2019bacterial,bhattacharjee2019confinement}. In the presence of the local nutrient gradient established through consumption, bacteria perform chemotaxis, biasing this random walk \cite{Bergbook}; the function $f(c)\equiv\log \left[\left(1+c/c_{-}\right)/\left(1+c/c_{+}\right)\right]$ describes the ability of the bacteria to logarithmically sense nutrient with characteristic concentrations $c_{-}$ and $c_{+}$ \cite{Cremer2019,fu2018spatial}, and the chemotactic coefficient $\chi$ describes their ability to then bias their motion in response
to the sensed nutrient gradient \cite{Keller1971a,fu2018spatial,Cremer2019}. The chemotactic velocity is thus given by $\bm{v}_{\text{ch}} \equiv \chi \bm{\nabla} f(c)$, where similar to $D_\text{b}$, the value of $\chi$ depends on both intrinsic cellular properties and pore-scale confinement \cite{Bhattacharjee2020}. Together, \cref{eq nutrient,eq bacteria-main-withgrowth} provide a continuum model of chemotactic migration that has thus far been successfully used to describe unperturbed \textit{E. coli} populations \cite{Keller1971a,fu2018spatial,Cremer2019,Bhattacharjee2020}.

\begin{figure*}
\centering
\includegraphics[width=0.75\textwidth]{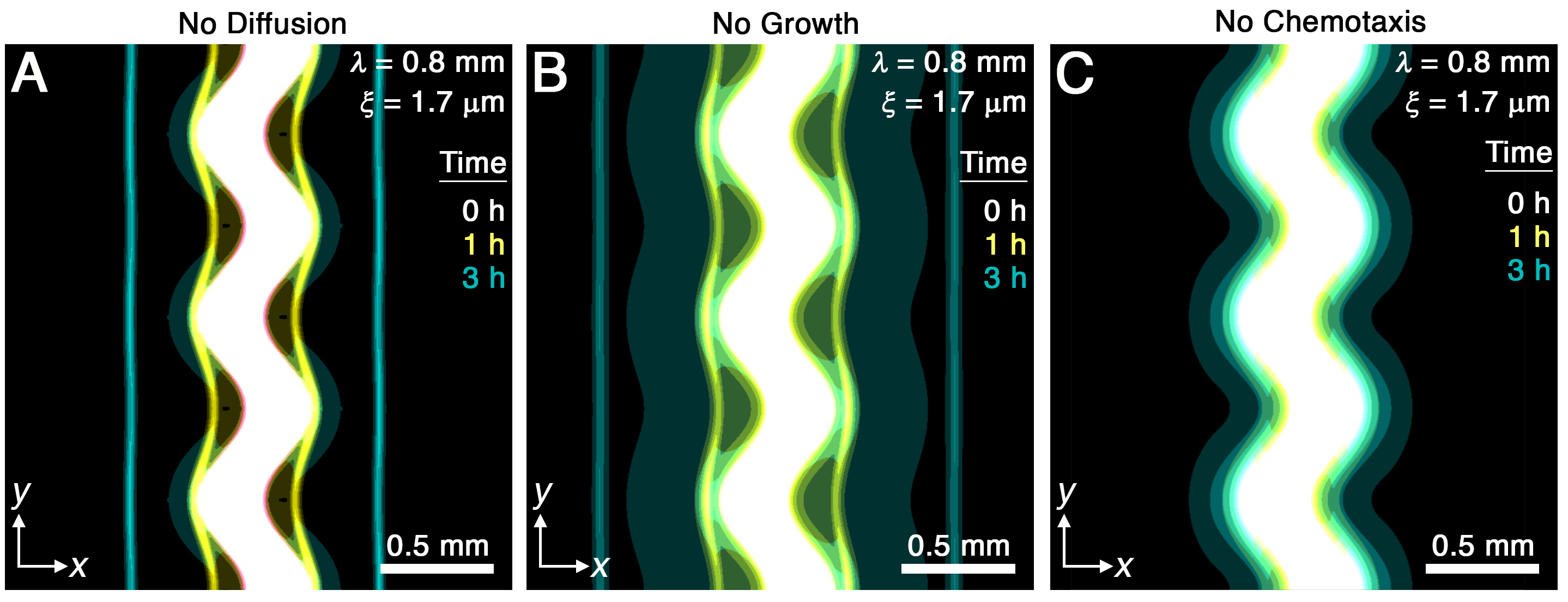}
  {\phantomsubcaption\label{Fig3a}}
  {\phantomsubcaption\label{Fig3b}}
  {\phantomsubcaption\label{Fig3c}}
\caption{\textbf{Chemotaxis is the primary driver of morphological smoothing.} Images show the same simulation as in \cref{Fig2a}, which serves as an exemplary case, but with either \subref*{Fig3a} diffusive cell motion, \subref*{Fig3b} cell proliferation, or \subref*{Fig3c} cell chemotaxis knocked out by setting the diffusivity $D_{b}$, proliferation rate $\gamma$, or chemotactic coefficient $\chi$ to zero, respectively. Simulated bacterial fronts lacking diffusion or proliferation still smooth, as shown in \subref*{Fig3a}-\subref*{Fig3b}, but simulated fronts lacking chemotaxis do not smooth, as shown in \subref*{Fig3c}---demonstrating that chemotaxis is necessary and sufficient for the observed morphological smoothing.}
\label{fig:fig3}
\end{figure*}

To simulate the chemotactic migration of perturbed populations, we numerically solve \cref{eq nutrient,eq bacteria-main-withgrowth} using undulated morphologies as initial conditions for $b$, similar to those explored in the experiments. The simulations employ values for all parameters
based on direct measurements, as detailed in the \hyperref[methods]{Materials and Methods}. Although we do not expect perfect quantitative agreement between the experiments and simulations due to their difference in dimensionality and the simplified treatment of cell-cell interactions, the simulated fronts form, collectively migrate, and smooth in a manner that is remarkably similar to the experiments. Three examples are shown in \crefrange{Fig2c}{Fig2e} (\hyperref[movies]{Movies S4 to S6}), corresponding to the experiments shown in \crefrange{Fig1c}{Fig1e} (\hyperref[movies]{Movies S1 to S3}). Similar to the experiments, the outer periphery of each population first spreads slowly, then spontaneously organizes into an outward-migrating front that eventually smooths. We again find that the normalized undulation amplitude decays exponentially over time, as shown in \cref{Fig2d}. As in the experiments, increasing the undulation wavelength $\lambda$ slows smoothing; compare \cref{Fig2b} to \cref{Fig2a}. Also as in the experiments, increasing the pore size $\xi$, which increases the migration parameters $D_\text{b}$ and $\chi$, greatly hastens smoothing; compare \cref{Fig2c} to \cref{Fig2a}. This variation of the smoothing time scale $\tau$ obtained from simulations with $\lambda$ and $\xi$ is summarized in \cref{Fig2e}. We observe the same behavior as in the experiments, with the absolute values of $\tau$ agreeing to within a factor of $\sim3$. This agreement confirms that the continuum Keller-Segel model recapitulates the essential spatio-temporal features of smoothing seen in the experiments.

\subsection*{Chemotaxis is the primary driver of front smoothing} The simulations provide a way to directly assess the relative importance of cellular diffusion, chemotaxis, and cell proliferation to front smoothing. To this end, we perform the same simulation as in \cref{Fig2a}, but with each of the corresponding three terms in \cref{eq bacteria-main-withgrowth} knocked out, and determine the resulting impact on collective migration. This procedure enables us to determine the factors necessary for smoothing. 
    
While diffusion typically causes spatial inhomogeneities to smooth out, we do not expect it to play an appreciable role in the front smoothing observed here: the characteristic time scale over which undulations of wavelength $\lambda\approx1$ mm diffusively smooth is $\sim \lambda^{2}/D_\text{b}\approx100$ to $700$ h, up to three orders of magnitude larger than the smoothing time $\tau$ measured in experiments and simulations. We therefore expect that the undirected motion of bacteria is much too slow to contribute to front smoothing. The simulations for $\lambda=0.8$ mm and $\xi=1.7~\mu$m confirm this expectation: setting $D_\text{b}=0$ yields fronts that still smooth over a time scale $\tau\sim1$ h similar to the full simulations (\cref{Fig3a}).

Another possible mechanism of front smoothing is differences in bacterial proliferation at different locations along the front periphery---for example, the front would smooth if cells in concave regions were able to proliferate faster than those in convex regions. However, differential proliferation typically destabilizes bacterial communities, as shown previously both experimentally and theoretically \cite{fujikawa,bonachela,nadell,farrell,Trinschek2018a,Allen2019}. Furthermore, even if proliferation were to help smooth the overall population, we again expect this hypothetical mechanism to be too slow to appreciably contribute: the shortest time scale over which cells all growing exponentially at a maximal rate $\gamma\sim1$ h$^{-1}$ spread over the length scale $A_{0}\approx300~\mu$m by growing end-to-end is $\gamma^{-1}\log_{2}\left(A_{0}/l_\text{cell}\right)\sim7$ h, where $l_\text{cell}\approx2~\mu$m is the cell body length. This time scale is over an order of magnitude larger than the $\tau$ measured in experiments and simulations. The simulations again confirm our expectation: setting $\gamma=0$ yields fronts that still smooth over a time scale $\tau\sim1$ h similar to the full simulations (\cref{Fig3b}).

These findings leave chemotaxis as the remaining possible mechanism of front smoothing. The simulations confirm this expectation: setting $\chi=0$ yields a population that slowly spreads via diffusion and proliferation, but that does not form collectively migrating fronts at all (\cref{Fig3c}). Therefore, chemotaxis is both necessary and sufficient for the observed front smoothing.

\begin{figure*}[t]
\centering
\includegraphics[width=0.75\textwidth]{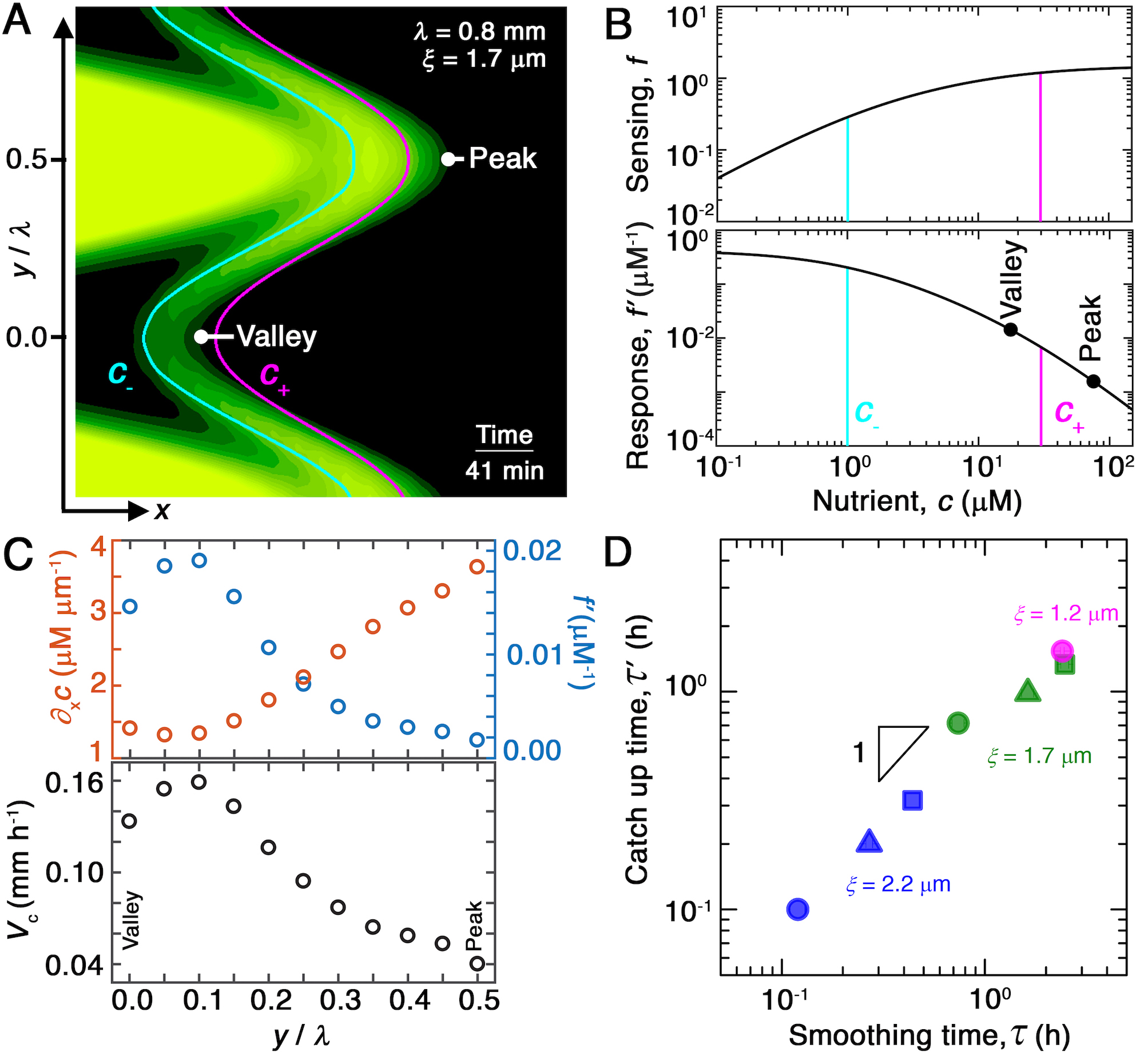}
  {\phantomsubcaption\label{Fig4a}}
  {\phantomsubcaption\label{Fig4b}}
  {\phantomsubcaption\label{Fig4c}}
  {\phantomsubcaption\label{Fig4d}}
\caption{\textbf{Chemotaxis alters the morphology of migrating bacterial fronts in two distinct ways.} \subref*{Fig4a} Magnified view of a migrating bacterial front from the simulation shown in \cref{Fig2a} at time $t=41$ min as a representative example. To illustrate the spatially-varying nutrient levels, we show the contours of constant nutrient concentration $c=c_{+}$ and $c=c_{-}$ in magenta and cyan, respectively; these represent characteristic upper and lower limits of sensing. The contours are spaced closer at the leading edge of the convex peak ($y/\lambda=0.5$) than the concave valley ($y/\lambda=0$), indicating that the magnitude of the local nutrient gradient is larger at peaks than at valleys. The nutrient concentration itself, which increases monotonically with increasing $x$, is also larger at the peak than at the valley. \subref*{Fig4b} Top and bottom panels show the variation of the nutrient sensing function $f(c)$ and chemotactic response function $f'(c)$, respectively, with nutrient concentration $c$. Because sensing saturates at high nutrient concentrations, chemotactic response is weaker at higher $c$ (peaks) than at lower $c$ (valleys). \subref*{Fig4c} Top panel shows the $x$ component of the nutrient gradient $\partial_{x}c$ (red, left axis) and the response function $f'$ (blue, right axis), and bottom panel shows the $x$ component of the chemotactic velocity $v_{\text{c},x}=\chi f' \partial_{x}c$ computed from these quantities, evaluated at different lateral positions $y$ along the leading edge of the front in \subref*{Fig4a}. While the driving force of chemotaxis represented by $\partial_{x}c$ is smaller at the valley, the chemotactic response $\chi f'$ is larger at the valley and dominates in setting $v_{\text{c},x}$: valleys move out faster than peaks, eventually catching up to them and smoothing out the undulations. \subref*{Fig4d} For all simulations (\cref{Fig2e}), the smoothing time $\tau$ determined by analyzing the decay of large-scale undulations (\cref{Fig2d}) is set by the time $\tau'$ needed for valleys to catch up to peaks estimated using their different $x$-component chemotactic velocities.}
\label{fig:fig4}
\end{figure*}

\subsection*{Distinct modes by which chemotaxis impacts front morphology} How exactly does chemotaxis smooth bacterial fronts? To address this question, we examine the spatially-varying chemotactic velocity $\bm{v}_{\text{c}} = \chi \bm{\nabla} f(c)$, which quantifies how rapidly different regions of the population migrate \textit{via} chemotaxis. To gain intuition for the determinants of $\bm{v}_{\text{c}}$, we recast this expression in terms of the nutrient gradient:
\begin{equation} \label{eq response}
    \bm{v}_{\text{c}} = {\underbrace{\chi f'(c)}_{\text{Response function}}}~{\underbrace{\bm{\nabla}c}_{\text{Forcing}}}.
\end{equation}
As in linear response theory, the chemotactic velocity can be viewed as the bacterial response to the driving force given by the nutrient gradient, $\bm{\nabla}c$, modulated by the chemotactic response function $\chi f'(c)$. Thus, variations in chemotactic velocity along the leading edge of the front, which specify how the overall front morphology evolves, are determined by the combined effect of variations in the nutrient gradient and the chemotactic response function. We therefore examine each of these modes by which chemotaxis influences front morphology in turn.

We first consider the nutrient gradient, which is the typical focus of chemotaxis studies. Our simulations, which numerically solve the coupled system of \cref{eq nutrient,eq bacteria-main-withgrowth}, directly yield the spatially-varying nutrient field $c$ and therefore $\bm{\nabla}c$. A snapshot from the representative example of \cref{Fig2a} is shown in \cref{Fig4a}, with the contours of $c=c_{-}$ and $c=c_{+}$ indicated by the cyan and magenta lines, respectively. The contours are spaced closer at the convex ``peaks'' (e.g., at $y/\lambda=0.5$) than at the concave ``valleys'' (e.g., at $y/\lambda=0$) along the leading edge of the front. Thus, the magnitude of the driving force given by $\bm{\nabla}c$ is larger at the peaks. We confirm this expectation by directly quantifying the nutrient gradient along the leading edge, focusing on the component $\partial_{x}c$ in the overall front propagation direction ($x$) for simplicity, as shown by the orange symbols in \cref{Fig4c}; as expected, this driving force is stronger at the peaks. This spatial variation in the driving force promotes faster outward chemotactic migration at the peaks than at the valleys, amplifying front undulations---in opposition to our observation that the migrating population self-smooths. Variations in the local nutrient gradient along the leading edge of the front do not contribute to smoothing; rather, they oppose it. 

We next turn to the chemotactic response function, which characterizes cellular signal transduction. Because $\chi$ is a constant for each porous medium \cite{Bhattacharjee2020}, spatial variations in the response function are set by variations in $f'(c)$. The sensing function $f(c)$ is plotted in the upper panel of \cref{Fig4b}. It varies linearly as $\sim c\left(1/c_{-}-1/c_{+}\right)$ for $c\ll c_{-}$ and saturates at $\log\left(c_{+}/c_{-}\right)$ for $c\gg c_{+}$; the characteristic concentrations $c_{-}$ and $c_{+}$ represent the dissociation constants of the nutrient for the inactive and active conformations of the cell-surface receptors, respectively \cite{Cremer2019,fu2018spatial,dufour,yang2015relation}. The response function $\chi f'(c)$ therefore decreases strongly as $c$ increases above $c_{+}$, which accordingly is often referred to as an upper limit of sensing (\cref{Fig4b}, lower panel). That is, because high nutrient concentrations saturate cell-surface receptors, the chemotactic response function decreases with nutrient concentration. Inspection of the nutrient field indicates that nutrient concentrations are larger at the peaks than at the valleys along the leading edge of the front (\cref{Fig4a}). Thus, the chemotactic response of cells is weaker at peaks than at valleys, as shown by the points in \cref{Fig4b}, yielding slower outward chemotactic migration at peaks than at valleys and thereby reducing the amplitude of front undulations. Variations in the chemotactic response along the leading edge of the front promote smoothing, unlike variations in the nutrient gradient. 

\subsection*{Spatial variations in chemotactic response\\drive morphological smoothing} We therefore hypothesize that the stabilizing effect of the chemotactic response (\cref{Fig4c}, blue) dominates over the destabilizing influence of the nutrient gradient (\cref{Fig4c}, red), leading to smoothing. Computation of the spatially-varying chemotactic velocity at the leading edge of the front using \cref{eq response}, focusing on the $x$ velocity component $v_{\text{c},x}\approx\chi f'\partial_{x}c$ for simplicity, supports this hypothesis: cells at concave regions migrate outward faster than those at convex regions (\cref{Fig4c}, lower panel). To further test this hypothesis, we assess the influence of varying $c_{+}$; we expect that reducing this upper limit weakens chemotactic response not just at the peaks, but also the valleys, thereby slowing smoothing. While tuning solely $c_{+}$ is challenging in the experiments, this can be readily done in the simulation---yielding slower smoothing, as expected (\cref{FigS1}). As a final test of our hypothesis, for each simulation shown in \cref{fig:fig2}, we determine the difference between the chemotactic velocities of the valleys and peaks, approximated by $\Delta v_{\text{c},x}\approx\chi [\left(f'\partial_{x}c\right)_{\text{valley}}-\left(f'\partial_{x}c\right)_{\text{peak}} ]$, as a function of time $\Delta t$. If smoothing is indeed due to variations of the chemotactic velocity along the leading edge, then the smoothing time $\tau$ determined by analyzing the decay of large-scale undulations (\cref{Fig2d,Fig2e}) should be approximately given by the time $\tau'$ at which valleys catch up to peaks, \textit{i.e.}, $\int_{0}^{\tau'}\Delta v_{\text{c},x}\;\dd\Delta t\approx A_{0}$. The $\tau'$ thus obtained is shown for all of our simulations of varying $\lambda$ and $\xi$ in \cref{Fig4d}. We find excellent agreement in all cases between $\tau'$ and $\tau$, shown on the vertical and horizontal axes respectively---confirming that smoothing is indeed determined by spatial variations in chemotactic velocity.

\section*{Discussion}
By combining experiments and simulations, this work elucidates a mechanism by which collectively migrating populations can smooth out large-scale perturbations in their overall morphology. We focus on the canonical example of chemotactic migration, in which coherent fronts of cells move in response to a self-generated nutrient gradient. Over the past half century, extensive studies have focused on the migration of unperturbed flat fronts \cite{adler1966effect,lauffenburger1991quantitative,Keller1971a,Cremer2019,fu2018spatial,saragosti2011directional,Bhattacharjee2020}; our work now demonstrates how perturbed fronts smooth out.  

The 3D printing platform provides a unique way to tune the shape of the initial perturbation, as well as the extent to which cellular migration is hindered. Our experiments using this approach reveal that the dynamics of smoothing are regulated by both the undulation wavelength and the ease with which cells migrate. The continuum simulations recapitulate the essential features of this behavior and shed light on the underlying mechanism; while studies of chemotaxis typically focus on the role of the nutrient gradient in driving cellular migration, our work highlights the distinct and pivotal role played by the chemotactic response function in regulating migration and large-scale population morphology. In particular, we find that even though cells in peaks of an undulated front experience a stronger driving force given by the local nutrient gradient, the higher nutrient levels they are exposed to saturate their cell-surface receptors, and hence they exhibit a weaker chemotactic response than cells in valleys. That is, while variations in the nutrient gradient along the leading edge of a front act to amplify undulations, variations in the ability of cells to sense and respond to this gradient dominate and instead smooth out the undulation. 

Our work thus reveals how chemotaxis in response to a self-generated nutrient gradient can enable a migrating population to withstand large-scale perturbations, providing a counterpoint to previous studies investigating the ability of perturbations to instead disrupt collective migration \cite{Sandor,Morin2017,Yllanes2017,Bera2020,Chepizhko2013,Chepizhko2013a,Chepizhko2015,Toner2018,Wong2014,Alert2020,Alert2019,Driscoll2017a,Doostmohammadi2016,Williamson2018,saverio,koch,kela1,kela2,BenAmar2016,BenAmar2016b,funaki,brenner,mimura,stark}. The chemotactic smoothing process is autonomous, arising without any external intervention; instead, it is a population-scale consequence of the limitations in cellular signal  transduction---motivating future studies of other population-scale effects that may emerge from individual behaviors. By demonstrating how chemotaxis drives smoothing, our work contributes a new factor to be considered in descriptions of morphogenesis, which thus far have focused on the role of other factors---such as differential proliferation, intercellular mechanics, substrate interactions, and osmotic stresses \cite{fujikawa,bonachela,nadell,farrell,Trinschek2018a,Allen2019,beroz,fei,yan,yan2,copenhagen,smith,ghosh}---in regulating the overall morphology of a bacterial population. Finally, because many other active systems such as other prokaryotes, cancer cells, white blood cells, amoeba, enzymes, chemically-sensitive colloidal microswimmers, and chemical robots \cite{iglesias,palagireview,granick,reichhardt2014,reichhardt2018,028bechinger,Konstantopoulos} also migrate \textit{via} chemotaxis, this mechanism of smoothing could broadly manifest in diverse forms of active matter.



\section*{Acknowledgments}
It is a pleasure to acknowledge Tommy Angelini for providing microgel polymers; Bob Austin for providing fluorescent \textit{E. coli}; and Stas Shvartsman, Howard Stone, Sankaran Sundaresan, and Ned Wingreen for stimulating discussions. This work was supported by NSF grant CBET-1941716, the Project X Innovation fund, a distinguished postdoctoral fellowship from the Andlinger Center for Energy and the Environment at Princeton University to T.B., the Eric and Wendy Schmidt Transformative Technology Fund at Princeton, the Princeton Catalysis Initiative, and in part by funding from the Princeton Center for Complex Materials, a Materials Research Science and Engineering Center supported by NSF grant DMR-2011750. This material is also based upon work supported by the National Science Foundation Graduate Research Fellowship Program (to J.A.O.) under Grant No. DGE-1656466. Any opinions, findings, and conclusions or recommendations expressed in this material are those of the authors and do not necessarily reflect the views of the National Science Foundation. R.A. acknowledges support from the Human Frontier Science Program (LT000475/2018-C).

\section*{Author contributions}

T.B. and S.S.D. designed the experiments; T.B. performed all experiments with assistance from J.A.O.; D.B.A., J.A.O., and S.S.D. designed the numerical simulations; D.B.A. performed all numerical simulations with assistance from J.A.O.; R.A. performed all theoretical calculations through discussions with S.S.D.; T.B., D.B.A., R.A., and S.S.D. analyzed the data; S.S.D. designed and supervised the overall project. All authors discussed the results and implications and wrote the manuscript.

\section*{Competing interests}
The experimental platform used to 3D print and image bacterial communities in this publication is the subject of a patent application filed by Princeton University on behalf of T.B. and S.S.D.

\section*{Data availability}
All data are available from the authors upon request.

\section*{Code availability}
All codes are available from the authors upon request.

\onecolumngrid

\clearpage

\twocolumngrid

\section*{Materials and Methods} \label{methods}

\subsection*{Preparing and characterizing porous media}We prepare 3D porous media by dispersing dry granules of crosslinked acrylic acid/alkyl acrylate copolymers (Carbomer 980, Ashland) in liquid EZ Rich, a defined rich medium for \textit{E. coli}. The components to prepare the EZ Rich are purchased from Teknova Inc., autoclaved prior to use, and are mixed following manufacturer directions; specifically, the liquid medium is an aqueous solution of 10X MOPS Mixture (M2101), 10X ACGU solution (M2103), 5X Supplement EZ solution (M2104), 20\% glucose solution (G0520), 0.132 M potassium phosphate dibasic solution (M2102), and ultrapure milli-Q water at volume fractions of 10\%, 10\%, 20\%, 1\%, 1\%, and 58\%, respectively. We ensure homogeneous dispersions of swollen hydrogel particles by mixing each dispersion for at least 2 h at 1600 rpm using magnetic stirring, and adjust the pH to 7.4 by adding 10 N NaOH to ensure optimal cell viability. The hydrogel granules swell considerably, resulting in a jammed medium made of $\sim5$ to $10~\mu$m diameter swollen hydrogel particles with $\sim20\%$ polydispersity and with an individual mesh size of $\sim40$ to $100$ nm, as we established previously \cite{bhattacharjee2019confinement}, which enables small molecules (e.g., amino acids, glucose, oxygen) to freely diffuse throughout the medium. 

Tuning the mass fraction of dispersed hydrogel particles enables the sizes of the pores between particles to be precisely tuned. We measure the smallest local pore dimension by tracking the diffusion of 200 nm-diameter fluorescent tracers through the pore space, as we detailed in a previous paper \cite{Bhattacharjee2020}. This previous paper shows the full pore size distributions thereby measured for porous media prepared in an identical manner to those used here; in this present paper, we only describe each medium using the mean pore size $\xi$, for simplicity.

\subsection*{3D printing bacterial populations}
Prior to each experiment, we prepare an overnight culture of \textit{E. coli} W3110 in LB media at 30$^{\circ}$C. We then incubate a 1\% solution of this culture in fresh LB media for 3 h until the optical density reaches $\sim0.6$, and then resuspend the cells in liquid EZ Rich to a concentration of $8.6\times10^{10}$ cells/mL. We then use this suspension as the inoculum that is 3D printed into a porous medium using a pulled glass capillary with a $\sim100$ to $200~\mu$m-wide opening as an injection nozzle. Each porous medium has a large volume of 4 mL and is confined in a transparent-walled glass-bottom petri dish 35 mm in diameter and 10 mm in height; in each experiment, the injection nozzle is mounted on a motorized translation stage that traces out a programmed two-dimensional undulating path within the porous medium, at least $\sim500$ to $1000$ $\mu$m away from any boundaries, at a constant speed of 1 mm/s. As the injection nozzle moves through the medium, it locally rearranges the hydrogel packing and gently extrudes the cell suspension into the interstitial space using a flow-controlled syringe pump at 50 $\mu$L/hr, which corresponds to a gentle shear rate of $\sim4$ to $36$ s$^{-1}$ at the tip of the injection nozzle. As the nozzle continues to move, the surrounding hydrogel particles rapidly densify around the newly-introduced cells, re-forming a jammed solid matrix \cite{bhattacharjee2018,bhattacharjee2015, bhattacharjee2016} that compresses the cellular suspension until the cells are close-packed to an approximate density of $0.95\times10^{12}$ cells/mL. This protocol thus results in a 3D-printed bacterial population having a defined initial amplitude and wavelength. Moreover, as we showed in our previous work \cite{Bhattacharjee2020}, this process does not appreciably alter the properties of the hydrogel packing and is sufficiently gentle to maintain the viability and motility of the cells. 

\subsection*{Imaging bacteria within porous media} 
Because the 3D-printed undulated cylinders of dense-packed cells are $\sim1$ cm long, each printing process requires $\sim10$ s. After 3D printing, the top surface of the porous medium is sealed with a thin layer of 1 to 2 mL of paraffin oil to minimize evaporation while allowing unimpeded oxygen diffusion. We then commence imaging within a few minutes after printing. Once an undulated population is 3D printed, it maintains its shape until cells start to move outward through the pore space. The time needed to print each cylinder is two orders of magnitude shorter than the duration between successive 3D confocal image stacks. Moreover, the 3D printing is fast enough to be considered as instantaneous when compared with the speed of bacterial migration. Thus, the imaging is sufficiently fast to capture the front propagation dynamics. To image how the distribution of cells evolves over time, we use a Nikon A1R+ inverted laser-scanning confocal microscope maintained at $30\pm1^{\circ}$C. In each experiment, we acquire vertical stacks of planar fluorescence images separated by $2.58~\mu$m along the vertical ($z$) direction, successively every $2$ to $30$ minutes for up to $20$ h. We then produce a maximum intensity projection from each stack at every time frame with the logarithm of fluorescent intensities displayed at every pixel; examples are shown in \cref{fig:fig1}.

\subsection*{Characterizing experimental front dynamics} 
We use each maximum intensity projection at each time point to manually measure the time-dependent amplitude ($A$) and radial location of the front ($R_f$) as defined in \cref{Fig1b}, identifying the edges of the front as the positions at which the fluorescent signal from cells matches the background noise. 

As we showed in our previous work \cite{Bhattacharjee2020}, due to the initially high cell density in the population, inter-cell collisions limit outward migration of the population; a coherent outward-propagating front only forms after at least $\sim1$ h. Here, we do not focus on these initial transient dynamics, but instead examine the long-time smoothing behavior of undulated fronts. We do this by tracking the decay of the time-dependent undulation amplitude over time, as shown in \cref{Fig1f}; we identify the time $t_0$ at which smoothing is initiated as the earliest time at which the error associated with an exponential fit to the decay of $A(t)$ is minimized. The initial value $A_{0}$ is then given by $A(t_0)$.

 \subsection*{Details of continuum model} To mathematically model the dynamics of bacterial fronts, we use a continuum description of chemotactic migration that we previously showed captures the essential dynamical features of flat fronts \cite{Bhattacharjee2020}. This model extends previous work on the classic Keller-Segel model \cite{fu2018spatial,saragosti2011directional,Cremer2019,croze2011migration,keller1975necessary,Keller1971a,keller1970initiation,odell1976traveling,lauffenburger1991quantitative,Seyrich2019} to the case of dense populations in porous media. In particular, we consider a 2D representation of the population in the $xy$ plane for simplicity and describe the evolution of the nutrient concentration $c(\bm{r},t)$ and number density of bacteria $b(\bm{r},t)$ using the coupled \cref{eq nutrient,eq bacteria-main-withgrowth}.
 
\textit{Nutrient diffusion and consumption.} The media used in our experiments have $L$-serine as the most abundant nutrient source and chemoattractant \cite{neidhardt1974culture}. \textit{E. coli} consume this amino acid first \cite{yang2015relation} and respond to it most strongly as a chemoattractant compared to other components of the media \cite{Wong2014,mesibov1972chemotaxis,adler1966effect,menolascina2017logarithmic}. Furthermore, the nutrient levels of our liquid medium are nearly two orders of magnitude larger than the levels under which \textit{E. coli} excrete appreciable amounts of their own chemoattractant \cite{budrene1991complex} and generate strikingly different front behavior \cite{budrene1991complex,budrene1995dynamics,Mittal2003} than those that arise in our experiments; however, the nutrient levels we use are sufficiently low to avoid toxicity associated with extremely large levels of $L$-serine \cite{neumann2014imprecision}. Thus, given all of these reasons, we focus on $L$-serine as the primary nutrient source and attractant, described by the scalar field $c(\bm{r},t)$. Equation \ref{eq nutrient} then relates changes in $c$ to nutrient diffusion and consumption by the bacteria. The nutrient diffusion coefficient $D_{\text{c}}=800~\mu$m$^2$/s is given by previous measurements in bulk liquid; we treat nutrient diffusion as being unhindered by the highly-swollen hydrogel matrix due to its large internal mesh size. The maximal consumption rate per cell $\kappa =1.6 \times 10^{-11}$ mM(cell/mL)$^{-1}$s$^{-1}$ is chosen based on previous measurements \cite{croze2011migration}, and $g(c)=c/\left(c+c_{\text{1/2}}\right)$ describes the influence of nutrient
availability relative to the characteristic concentration $c_{\text{1/2}}=1~\mu$M through Michaelis-Menten kinetics, as established previously \cite{Cremer2019,croze2011migration,monod1949growth,woodward1995spatio,shehata1971effect}. These values yield simulated fronts that we have previously validated against experiments in porous media for the unperturbed case \cite{Bhattacharjee2020}.

\textit{Bacterial diffusion and chemotaxis.} The bacterial flux $\bm{J}_{\text{b}}$ as included in \cref{eq bacteria-main-withgrowth} arises from the undirected and directed motion of cells, i.e., diffusion $-D_{\text{b}}\bm{\nabla} b$ and chemotaxis $b \chi \bm{\nabla} f(c)$, respectively. The value of the active cellular diffusion coefficient $D_\text{b}$ decreases with increasing pore-scale confinement \cite{bhattacharjee2019bacterial,bhattacharjee2019confinement}; as validated in our previous work \cite{bhattacharjee2019bacterial} for porous media identical to those used here, $D_{\text{b}}=2.32$, $0.93$, and $0.42~\mu$m$^2$/s for porous media with $\xi=2.2$, $1.7$, and $1.2~\mu$m, respectively. We describe cellular chemotaxis using the sensing function $f(c)\equiv\log \left[\left(1+c/c_{-}\right)/\left(1+c/c_{+}\right)\right]$ and the chemotactic coefficient $\chi$, as established previously \cite{fu2018spatial,Cremer2019}. The characteristic concentrations $c_{-}=1~\mu$M and $c_{+}=30~\mu$M represent the dissociation constants of the nutrient for the inactive and active conformations of the cell-surface receptors, respectively \cite{Cremer2019,fu2018spatial,dufour,yang2015relation,sourjik2012responding,shimizu2010modular,tu2008modeling,Kalinin2009,shoval2010fold,lazova2011response,celani2011molecular}. Similar to $D_\text{b}$, the value of the active chemotactic coefficient $\chi$ decreases with increasing pore-scale confinement \cite{Bhattacharjee2020}; by matching the long-time front propagation speed in our simulations with the experiments, we obtain $\chi = 145$, $9$, and $5~\mu$m$^2$/s for porous media with $\xi=2.2$, $1.7$, and $1.2~\mu$m, respectively. Although heterogeneity in $D_\text{b}$ and $\chi$ may be present within each population itself \cite{fu2018spatial}, we focus our analysis on the influence of pore size by assuming a constant value of both for each simulation. Finally, we note that the motility parameters $D_{\text{b}}$ and $\chi$ reflect the ability of cells to move through the pore space \textit{via} an unbiased or biased random walk with mean step length $l$ whose value depends on pore-scale confinement and possible cell-cell collisions in the pore space. For the case of sufficiently dilute cells in porous media, $l$ is set by the geometry of the pore space, as we previously established \cite{bhattacharjee2019bacterial,bhattacharjee2019confinement}; in particular, $l\approx l_{\text{c}}$, the mean length of chords, or straight paths that fit in the pore space \cite{torquato1993chord}. However, when the cells are sufficiently dense, as arises in the experiments explored here, cell-cell collisions truncate $l$. We model this by considering the mean separation between cells $l_{\text{cell}}\approx\left(\frac{3f}{4 \pi b}\right)^{1/3}-d$, where $f$ is the volume fraction of the pore space between hydrogel particles, $b$ is the local bacterial number density, and $d \approx 1~\mu$m is the characteristic size of a cell; for simplicity, when $l_{\text{cell}}< l_{\text{c}}$, we assume that cell-cell collisions truncate the mean step length $l$ and set its value to $l_{\text{cell}}$. That is, wherever $b$ is so large that $l_{\text{cell}}< l_{\text{c}}$, we multiply the values of both $D_{\text{b}}$ and $\chi$ used in \cref{eq bacteria-main-withgrowth} by the correction factor $(l_{\text{cell}}/l_{\text{c}})^2$ that accounts for the truncated $l$ due to cell-cell collisions. Moreover, wherever $b$ is even so large that this correction factor is less than zero---i.e. cells are jammed---we set both $D_{\text{b}}$ and $\chi$ to zero. Based on our experimental characterization of pore space structure \cite{bhattacharjee2019confinement} we use $f=0.36$, $0.17$, and $0.04$, and $l_{\text{c}}=4.6$, $3.1$, and $2.4~\mu$m, for porous media with $\xi=2.2$, $1.7$, and $1.2~\mu$m, respectively.

\textit{Bacterial proliferation.} Changes in $b$ can also arise from net cell proliferation, as described in \cref{eq bacteria-main-withgrowth}. In particular, we describe net cell proliferation with the maximal rate per cell $\gamma$ multiplied by the Michaelis-Menten function $g(c)$ that again describes describes the influence of nutrient availability i.e. it quantifies the reduction in proliferation rate when nutrient is sparse. We directly measured $\gamma \equiv \ln{2}/\tau_2$ previously, where $\tau_2 = 60$ min is the mean cell division time in a porous medium for our experimental conditions. We note that because $c$ and $b$ are coupled in our model, we do not require an additional “carrying capacity” of the population to be included, as is often done \cite{Cremer2019,croze2011migration}; we track nutrient deprivation directly through the radially-symmetric nutrient field $c(\bm{r},t)$.

\subsection*{Implementation of numerical simulations} While the experimental geometry is three dimensional, in previous work \cite{Bhattacharjee2020}, we found that radial and out of plane effects do not need to be considered to capture the essential features of bacterial front formation and migration. Thus, for simplicity, we use a 2D representation. In the $x$ direction (coordinates defined in \cref{fig:fig2,fig:fig4}), no flux boundary conditions are used at the walls of the simulated region for both field variables $b$ and $c$. In the $y$ direction, no flux boundary conditions are used after one wavelength of the undulation, peak to peak, which comprises a single repeatable unit. The initial cylindrical distribution of cells 3D printed in the experiments has a diameter of $\sim100~\mu$m; so, in the $x$ dimension of the numerical simulations, we use a Gaussian with a $100~\mu$m full width at half maximum for the initial bacteria distribution $b(x,t=0)$, with a peak value that matches the 3D printed cell density in the experiments, $0.95 \times 10^{12}$ cells/mL. We vary the center $x$ position of the Gaussian distribution sinusoidally along $y$ to reproduce a given experimental wavelength and amplitude. Experimental wavelengths were measured directly from confocal images and rounded to the nearest $10~\mu$m. The initial condition of nutrient is $c=10$ mM everywhere, characteristic of the liquid media used in the experiments. The initial nutrient concentration is likely lower within the experimental population initially due to nutrient consumption during the 3D printing process; however, we expect this discrepancy to play a negligible role as nutrient deprivation occurs rapidly in the simulations.

As previously detailed \cite{Bhattacharjee2020}, while the periphery of a 3D printed bacterial population forms a propagating front, cells in the inner region remain fixed and eventually lose fluorescence because they are nutrient-limited. Specifically, the fluorescence intensity of this fixed inner population remains constant over an initial duration $\tau_{\text{delay}} = 2$ h, and then exponentially decreases with a decay time scale $\tau_{\text{starve}} = 29.7$ min. To facilitate comparison to the experiments, our simulations incorporate this feature to represent the cellular signal, which is the analog of the fluorescence measured in experiments, in \cref{fig:fig2,fig:fig4}. We do this by multiplying the cellular density obtained by solving \cref{eq bacteria-main-withgrowth} by a correction factor that incorporates the history of nutrient depletion. Specifically, wherever $c(\bm{r}',t')$ drops below a threshold value, for all times $t>t'+\tau_{\text{delay}}$, we multiply the cellular density $b(\bm{r}',t)$ by $e^{-(t-t')/\tau_{\text{starve}}} $, where $t'$ is the time at which the position $\bm{r}'$ became nutrient-depleted. 

To numerically solve the continuum model, we use an Adams-Bashforth-Moulton predictor corrector method \cite{Seyrich2019} where the order of the predictor and corrector are 3 and 2, respectively. Since the predictor corrector method requires past time points to inform future steps, the starting time points must be found with another method; we choose the Shanks starter of order 6 \cite{shanks1966solutions}. For the first and second derivatives in space, we use finite difference equations with central difference forms in 2D. Time steps of the simulations are $0.01$ s and spatial resolution is $10~\mu$m. Because the experimental chambers are $3.5$ cm in diameter, we use a distance of $3.5\times 10^4~\mu$m for the size of the entire simulated system in the $x$ direction with the cells initially situated in the center. Our previous work \cite{Bhattacharjee2020} demonstrated that the choice of discretization does not appreciably influence the results in numerical simulations of flat fronts; furthermore, our new results for the simulations performed here (\cref{FigS2}) indicate that our choice of discretization used is sufficiently finely-resolved such that the results in numerical simulations of undulated fronts are not appreciably influenced by discretization.

\bigskip

\subsection*{Characterizing simulated front dynamics} For the analysis shown in \cref{fig:fig2}, the leading edge is defined as the locus of positions at which $b$ falls below a threshold value equal to $10^{-4}$ times the maximum cell density of the initial bacterial distribution, as in \cite{Bhattacharjee2020}. For the analysis shown in \cref{fig:fig4}, to more accurately track the leading edge of the front, we define it as the locus of positions at which $b$ falls below a threshold value specific to each condition tested; the threshold is $0.003$ cells per $\mu$m$^3$ for the prototypical case of $\xi=1.7~\mu$m and $\lambda=0.8$ mm shown in \crefrange{Fig4a}{Fig4c}, as well as all simulations for $\xi=2.2~\mu$m; $0.002$ cells per $\mu$m$^3$ for simulations for $\xi=1.7~\mu$m and $\lambda=2.0$ and $3.2$ mm; and $0.001$ cells per $\mu$m$^3$ for simulations for $\xi=1.2~\mu$m and $\lambda=0.8$ mm.

\onecolumngrid 

\clearpage

\onecolumngrid
\begin{center}
\textbf{\large Supplementary Information}\\
\end{center}

\setcounter{equation}{0}
\setcounter{figure}{0}
\renewcommand{\theequation}{S\arabic{equation}}
\renewcommand{\thefigure}{S\arabic{figure}}


\section*{Supplementary Figures}

\begin{figure*}[!htb]
\centering
\includegraphics[width=0.3\textwidth]{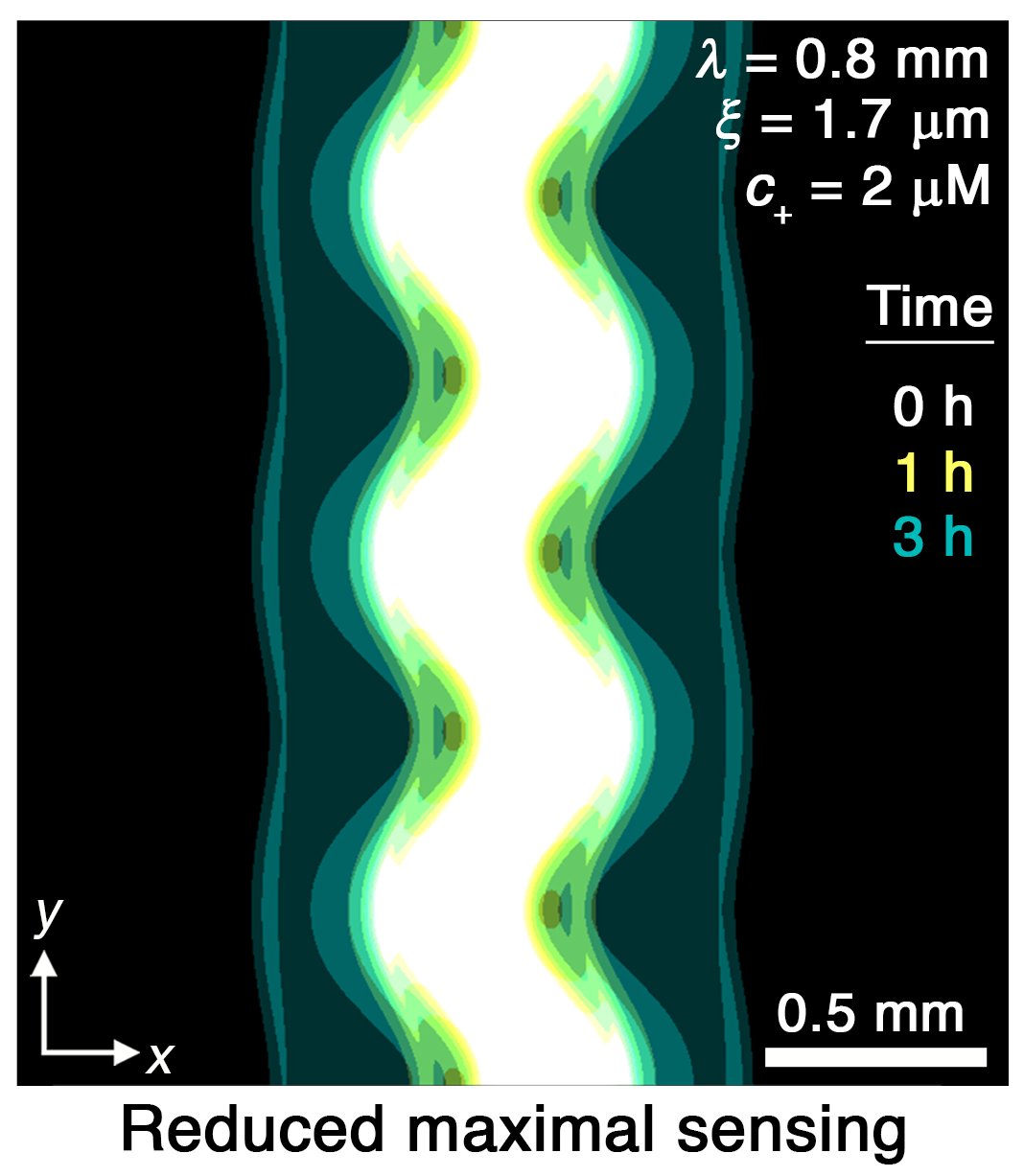}
\caption{\textbf{Effect of reduced sensing.} To investigate the influence of varying the upper limit of sensing $c_{+}$, we repeat the simulation for the prototypical case of $\xi=1.7~\mu$m and $\lambda=0.8$ mm but with $c_{+}$ lowered by a factor of $15$. Consistent with our expectation, we find that reducing this upper limit weakens chemotactic response not just at the peaks, but also the valleys, thereby slowing smoothing. Image is presented as in Fig. 2A.}
\label{FigS1}
\end{figure*}

\begin{figure*}[!htb]
\centering
\includegraphics[width=0.3\textwidth]{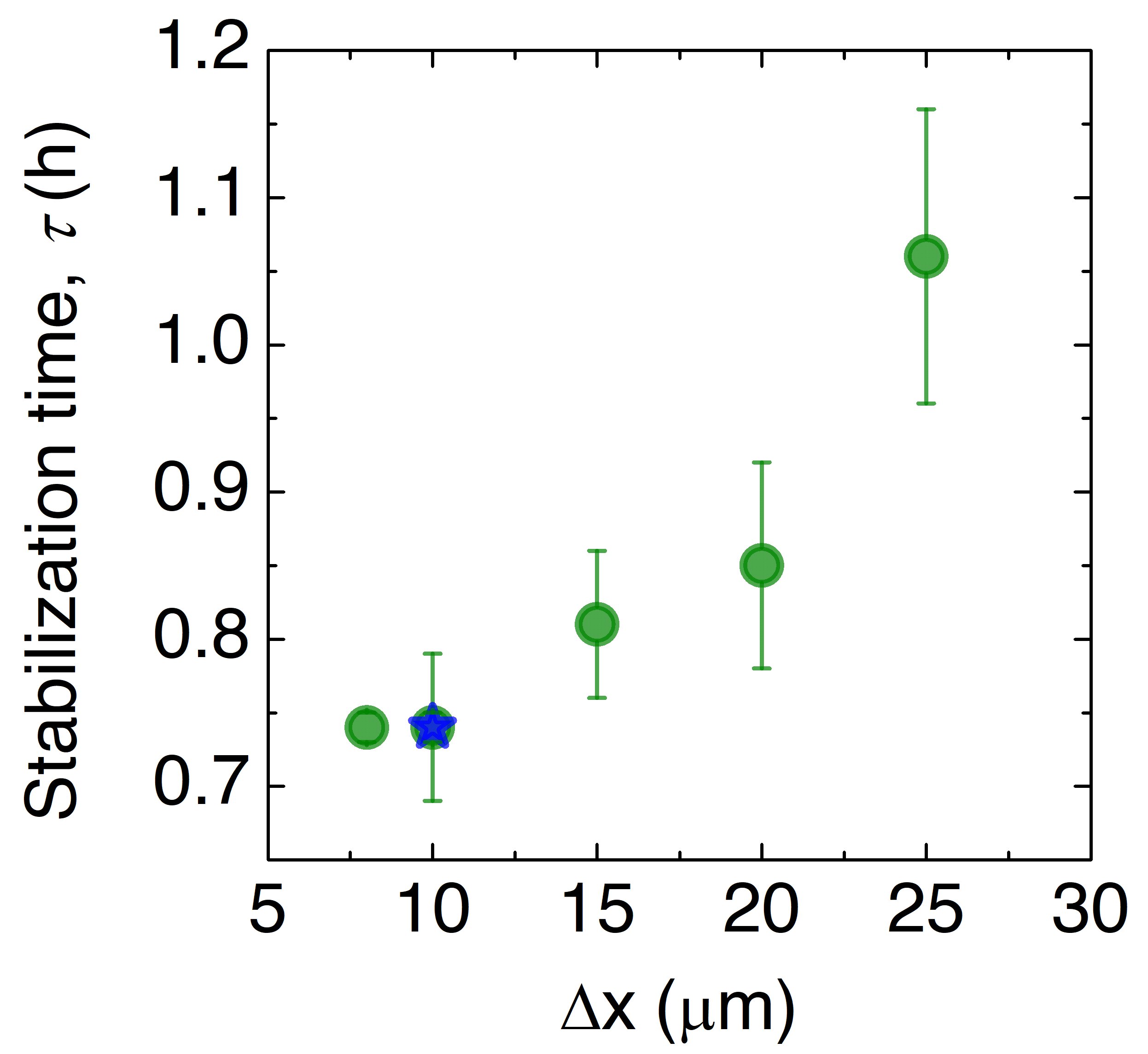}
\caption{\textbf{Convergence of the numerical simulations.} To assess the influence of discretization, we repeat the simulation for the prototypical case of $\xi=1.7~\mu$m and $\lambda=0.8$ mm with different choices of the spatial discretization $\Delta x$ and measure the smoothing time $\tau$. In all cases we find qualitatively similar results, although the dynamics vary; however, as shown by the green data points, the dynamics do not appreciably change for discretization smaller than $\approx10~\mu$m, which is the value used in the main text simulations, as indicated by the blue star.}
\label{FigS2}
\end{figure*}

\clearpage

\section*{Supplementary Movies} \label{movies}

\noindent\textbf{Movie S1: Experiment probing chemotactic smoothing for $\lambda=0.8$ mm, $\xi=1.7~\mu$m.} Movie shows the maximum intensity fluorescence projection (bottom up view) of migration from a 3D-printed undulated cylinder of close-packed \textit{E. coli}. The cells collectively migrate outward in a front that autonomously smooths out the large-scale undulations as it continues to propagate.

\bigskip

\noindent\textbf{Movie S2: Experiment probing chemotactic smoothing for $\lambda=3.4$ mm, $\xi=1.7~\mu$m.} Movie shows the maximum intensity fluorescence projection (bottom up view) of migration from a 3D-printed undulated cylinder of close-packed \textit{E. coli}.

\bigskip

\noindent\textbf{Movie S3: Experiment probing chemotactic smoothing for $\lambda=0.8$ mm, $\xi=2.2~\mu$m.} Movie shows the maximum intensity fluorescence projection (bottom up view) of migration from a 3D-printed undulated cylinder of close-packed \textit{E. coli}.

\bigskip

\noindent\textbf{Movie S4: Simulation probing chemotactic smoothing for $\lambda=0.8$ mm, $\xi=1.7~\mu$m.} Movie shows the calculated cellular fluorescence signal of cells migrating from an undulated stripe of close-packed \textit{E. coli} similar to \hyperref[movies]{Movie S1}. As in the experiments, the cells collectively migrate outward in a front that autonomously smooths out the large-scale undulations as it continues to propagate.

\bigskip

\noindent\textbf{Movie S5: Simulation probing chemotactic smoothing for $\lambda=3.4$ mm, $\xi=1.7~\mu$m.} Movie shows the calculated cellular signal fluorescence signal of cells migrating from an undulated stripe of close-packed \textit{E. coli} similar to \hyperref[movies]{Movie S2}.

\bigskip

\noindent\textbf{Movie S6: Simulation probing chemotactic smoothing for $\lambda=0.8$ mm, $\xi=2.2~\mu$m.} Movie shows the calculated cellular fluorescence signal of cells migrating from an undulated stripe of close-packed \textit{E. coli} similar to \hyperref[movies]{Movie S3}.

\end{document}